# Practical Portfolio Optimization with Metaheuristics: Pre-assignment Constraint and Margin Trading

Hang Kin Poon[1][0009-0007-4471-8902]

[1]Hong Kong Metropolitan University
`s1375690@live.hkmu.edu.hk`

**Abstract.** Portfolio optimization is a critical area in finance, aiming to maximize returns while minimizing risk. Metaheuristic algorithms were shown to solve complex optimization problems efficiently, with Genetic Algorithms and Particle Swarm Optimization being among the most popular methods. This paper introduces an innovative approach to portfolio optimization that incorporates pre-assignment to limit the search space for investor preferences and better results. Additionally, taking margin trading strategies in account and using a rare performance ratio to evaluate portfolio efficiency. Through an illustrative example, this paper demonstrates that the metaheuristic-based methodology yields superior risk-adjusted returns compared to traditional benchmarks. The results highlight the potential of metaheuristics with help of assets filtering in enhancing portfolio performance in terms of risk adjusted return.

**Keywords:** Metaheuristics, Evolutionary algorithms, Swarm intelligence, Portfolio optimization, Investment, Margin trading, MAR ratio.

## 1 Introduction

Portfolio optimization is a fundamental practice in finance, aimed at allocating assets to maximize returns while minimizing risk. This process involves constructing portfolios that optimize risk-adjusted returns, reflecting the return on investment relative to the risk taken. Risk-adjusted return measures how profitable portfolios are in terms of drawdown. However, traditional methods of portfolio optimization often struggle with the complexity and constraints of real-world financial markets.

Metaheuristic approaches have emerged as powerful tools for addressing these challenges. Unlike traditional optimization techniques, metaheuristics family which includes Particle Swarm Optimization (PSO) and Genetic Algorithms (GA) can handle complex, nonlinear, and constrained optimization problems [1]. These algorithms, modeled after natural processes, effectively navigate vast and intricate search spaces to uncover near-optimal solutions.

One key aspect of applying metaheuristics to portfolio optimization is pre-screening the search space. By incorporating pre-assignment constraints, investors can narrow down the search space to reflect real-world limitations and preferences. This pre-screening process ensures that the optimization algorithm focuses on feasible and relevant solutions, enhancing both the efficiency and practicality of the portfolio construction process.

Margin trading offers the potential for amplified returns by allowing investors to borrow funds to purchase assets. However, it also introduces significant risks, particularly the risk of a margin call. A margin call is triggered when the portfolio's value drops below a specified level, known as the maintenance margin, due to market fluctuations. When this occurs, the investor must deposit additional funds to maintain their position; otherwise, the broker may liquidate assets to cover the deficit. This situation can amplify losses and result in forced selling at disadvantageous prices.

To handle the risks associated with margin trading, it's important to evaluate the portfolio's maximum drawdown, which measures the largest drop in value from a peak to a trough before a new peak is achieved. By understanding and managing maximum drawdown, investors can better prepare for potential margin calls and implement strategies to prevent forced liquidation.

This paper aims to present a comprehensive framework for portfolio optimization using metaheuristic approaches, focusing on pre-screening of the search space and the importance of analyzing maximum drawdown in the context of margin trading. Through empirical analysis and illustrative examples, this paper compares the performance of the proposed framework with traditional benchmarks, demonstrating its potential to provide investors with more robust and adaptable portfolios in dynamic market conditions. Section 2 to 6 provides literature review and backgrounds necessary for research and experiment.

## 2     The Single-Objective Portfolio Optimization Problem

Portfolio optimization with one object focuses on optimizing one primary goal, typically maximizing return or minimizing risk. This approach simplifies the optimization process by concentrating on a single criterion, making it computationally efficient and easier to implement compared to multi-objective optimization.

One of the most straightforward applications of single-objective optimization is maximizing the expected return of a portfolio. In this scenario, the objective function is designed to select assets and their respective weights to achieve the highest possible return, given certain constraints such as budget limitations or sector allocations. While this approach can lead to high returns, it often comes with increased risk, as higher returns are typically associated with higher volatility.

An alternative single-objective approach is to minimize the risk of a portfolio, often measured by variance or standard deviation of returns [2]. This approach is particularly advantageous for risk-averse investors who focus more on safeguarding their capital than pursuing high-yield opportunities. By emphasizing risk reduction, investors can build portfolios that are more stable and resilient to market volatility. However, this focus on safety may result in lower expected returns, as less risky assets typically provide lower yields.

## 3 The Multi-objective Portfolio Optimization Problem

Multi-objective portfolio optimization extends the traditional single-objective approach by simultaneously addressing multiple, often conflicting objectives [3]. This methodology is crucial, which achieve a balance between competing factors such as maximizing returns and minimizing risks, among others. By incorporating multiple objectives, this approach provides a more comprehensive framework for decision-making in complex financial environments.

The general formulation of a multi-objective optimization problem can be expressed as follows:

$$\text{Minimize} \quad f(x) = (f_1(x), f_2(x), \dots, f_k(x))$$

$$\begin{aligned}
\text{subject to} \quad & g_m(x) \leq 0, m = 1, \dots, n_g \\
& h(x) = 0, m = n_g + 1, \dots, n_g + n_h \\
& x \in \Omega
\end{aligned} \quad (1)$$

Here, $x$ represents an n-dimensional decision variable vector within the search space $\Omega$. The functions $f_1(x), f_2(x), \dots, f_k(x)$ denote the $k$ objectives to be minimized, while $g_m(x)$ and $h(x)$ represent the inequality and equality constraints, respectively. The parameters $n_g$ and $n_h$ correspond to the number of inequality and equality constraints [4][5][2].

The concept of Pareto optimality is central to multi-objective optimization [4]. The Pareto front consists of the best possible solutions where improving one objective would necessitate compromising another. Investors can select a portfolio from the Pareto front based on their specific preferences and risk tolerance, ensuring a tailored and optimized investment strategy.

The most common application of multi-objective optimization in portfolio management is the simultaneous maximization of expected return and minimization of risk. Unlike single-objective optimization, which targets a single main goal, multi-objective optimization aims to discover a set of Pareto-optimal solutions. These solutions represent portfolios where improving any one objective would inevitably detract from another. This approach provides investors with a range of optimal portfolios, each offering a different trade-off between return and risk.

Beyond return and risk, multi-objective optimization can incorporate more various objectives to reflect the complexities of real-world investing. These may include Liquidity that ensures the portfolio maintains a certain level of liquidity to meet short-term cash needs or to manage risk. Diversification which maximizes the diversity of assets within the portfolio to reduce unsystematic risk. Transaction Costs that minimizes the costs associated with buying and selling assets to enhance overall portfolio performance.

Similar to single-objective optimization, multi-objective optimization can incorporate various constraints to reflect real-world limitations and investor preferences. These constraints may include budget limitations, sector and asset class limits, and regulatory requirements. The inclusion of constraints ensures that the optimization process yields practical and feasible solutions tailored to the investor's specific needs.

Metaheuristic algorithms, such as Genetic Algorithms (GA) and Particle Swarm Optimization (PSO), are well-suited for multi-objective optimization problems due to their ability to handle complex, nonlinear, and constrained search spaces [1]. These algorithms can efficiently explore the Pareto front, identifying a set of non-dominated solutions that represent optimal trade-offs between the multiple objectives [6]. Their flexibility and adaptability make them powerful tools for optimizing portfolios in dynamic and uncertain market conditions.

The primary advantage of multi-objective optimization is its ability to provide a more holistic view of portfolio performance, considering multiple important factors simultaneously. This method enables investors to make better-informed decisions by balancing multiple objectives in line with their preferences and risk tolerance. However, multi-objective optimization is more computationally intensive and complex than single-objective optimization, requiring sophisticated algorithms and careful interpretation of results.

In conclusion, multi-objective portfolio optimization offers a comprehensive and flexible approach to portfolio management, enabling investors to balance multiple objectives and constraints. By leveraging advanced metaheuristic algorithms, investors can achieve more robust and adaptable portfolios that better align with their financial goals and risk tolerance in dynamic market conditions.

## 4      Particle Swarm Optimization (PSO)

Particle Swarm Optimization (PSO) is in the family of metaheuristic algorithm and trying to simulate social behavior like fish school or bird flock [7]. It is getting significant attention in various optimization fields, including portfolio management, due to its simplicity and efficiency.

PSO works with a group of potential solutions, known as particles, that navigate the search space to identify optimal solutions. Each particle embodies a possible solution to the optimization problem and possesses two key attributes: Position that represents a specific solution in the search space and Velocity that determines the direction and speed of the particle's movement.

The positions of particles are adjusted by the experience of their neighbors and by their own experience, mimicking the social interaction and information sharing observed in natural swarms. The search space is explored and optimal is found by the collective behavior of PSO [8].

In the context of portfolio optimization, PSO can be employed to look for portfolio combination that balances multiple objectives. Each portfolio is a particle in the swarm, with its position encoding the weights of different assets [9]. The fitness function evaluates the performance of each portfolio based on the specified objectives, such as expected return, risk (e.g., variance or standard deviation), and other relevant criteria.

PSO offers several advantages for portfolio optimization. PSO is easy to implement and computationally efficient, making it suitable for large-scale optimization problems.

PSO can handle complex, nonlinear, and high-dimensional search spaces, making it well-suited for real-world portfolio optimization problems with numerous assets and constraints [10]. PSO can be easily adapted to incorporate various objectives and constraints, allowing for customized optimization strategies tailored to specific investor preferences and market conditions.

Despite its advantages, PSO also faces challenges and limitations. PSO's effectiveness is highly dependent on the selection of parameters like inertia weight and acceleration coefficients. Fine-tuning these parameters is crucial for achieving optimal results. PSO may converge prematurely to suboptimal solutions, particularly in complex search spaces with many local optima. Techniques such as dynamic parameter adjustment and hybridization with other algorithms can help mitigate this issue.

Particle Swarm Optimization (PSO) is a versatile and efficient metaheuristic algorithm for portfolio optimization [11]. Its ability to handle complex, nonlinear, and high-dimensional search spaces makes it a valuable tool for constructing optimal portfolios that balance return, risk, and other relevant objectives [10]. By leveraging the principles of swarm intelligence, PSO enables investors to navigate the intricacies of financial markets and achieve more robust and adaptable investment strategies.

## 5 Pre-assignment constraint

Pre-assignment constraints play a crucial role in portfolio optimization by imposing specific conditions on the allocation of assets before the optimization process begins. These constraints reflect real-world limitations, investor preferences, and regulatory requirements.

Pre-assignment constraints are rules or conditions that dictate certain aspects of the portfolio's structure or composition prior to optimization. They are used to incorporate specific requirements or preferences into the portfolio construction process, guiding the optimization algorithm towards feasible and relevant solutions. By setting pre-assignment constraints, investors can tailor the optimization process to better reflect their unique needs and market conditions.

Pre-assignment constraints can take various forms, depending on the investor's objectives and the market environment. Some common types of pre-assignment constraints include asset allocation limits which specify minimum or maximum percentages for certain asset classes, sectors, or individual securities within the portfolio. These limits help manage risk exposure and ensure diversification. Investment Exclusions restrict the inclusion of certain assets or sectors based on ethical considerations, risk aversion, or other criteria. Diversification Requirements ensure that the portfolio is sufficiently diversified by limiting the concentration of investments in any single asset or sector. This helps mitigate unsystematic risk and enhances the portfolio's resilience to market fluctuations. Liquidity Constraints set aside a portion of the portfolio in highly liquid assets to meet short-term cash needs or to manage risk. This ensures that the investor has access to funds when needed, without having to liquidate long-term investments. Risk Tolerance imposes constraints based on the investor's risk tolerance, such as limiting exposure to high-volatility assets. This helps align the portfolio with the investor's risk profile and financial goals.

Incorporating pre-assignment constraints into the portfolio optimization process offers several benefits. Pre-assignment constraints ensure that the resulting portfolio is practical and implementable, reflecting real-world limitations and investor preferences. By imposing constraints on asset allocation, diversification, and risk exposure, investors can better manage portfolio risk and enhance resilience to market fluctuations. Pre-assignment constraints allow investors to tailor the optimization process to their unique needs and circumstances, resulting in a more personalized and effective investment strategy.

While pre-assignment constraints offer numerous benefits, they also present challenges and considerations. Imposing constraints may limit the optimization algorithm's ability to find the global optimum, requiring investors to balance the benefits of constraints with potential trade-offs in performance. Pre-assignment constraints may need to be dynamically adjusted over time to reflect changing market conditions and investor preferences.

## 6    Margin trading and MAR ratio

Margin trading allows investors to increase their purchasing power by borrowing funds from a broker to invest in securities. While this can amplify potential gains, it also introduces significant risks, making it a double-edged sword in portfolio management. Investors must thoroughly understand the mechanics and implications of margin trading to effectively enhance portfolio performance while managing associated risks.

Margin trading involves using borrowed funds to control a larger position than would be possible with the investor's own capital alone. The process may involve:
1. Initial Margin: The investor deposits an initial amount, known as the initial margin, which represents a percentage of the total value of the securities purchased. This margin serves as collateral for the loan provided by the broker.
2. Leverage: The broker lends additional funds to the investor, allowing them to purchase more securities than they could with their own capital. Leverage amplifies both potential gains and losses, as the investor's exposure to market movements is increased.
3. Maintenance Margin: The investor must maintain a minimum amount of equity in the account, known as the maintenance margin. If the value of the securities falls below this threshold, the investor may face a margin call.
4. Margin Call: When a portfolio's value drops below the maintenance margin, the broker will issue a margin call, prompting the investor to add more funds to satisfy the margin requirement. If the investor fails to meet this call, the broker may sell off assets in the portfolio to address the deficit, potentially resulting in forced sales at disadvantageous prices.

As margin trading increases the portfolio's sensitivity to market volatility, making it more susceptible to price fluctuations and potential losses. Assessing the maximum drawdown of the portfolio is crucial in margin trading. A maximum drawdown (MDD) represents the largest loss observed in a portfolio from its highest point (peak) to its lowest point (trough) before reaching a new peak. As it measures the worst condition, it is particularly useful to avoid portfolios that may get investors margin called.

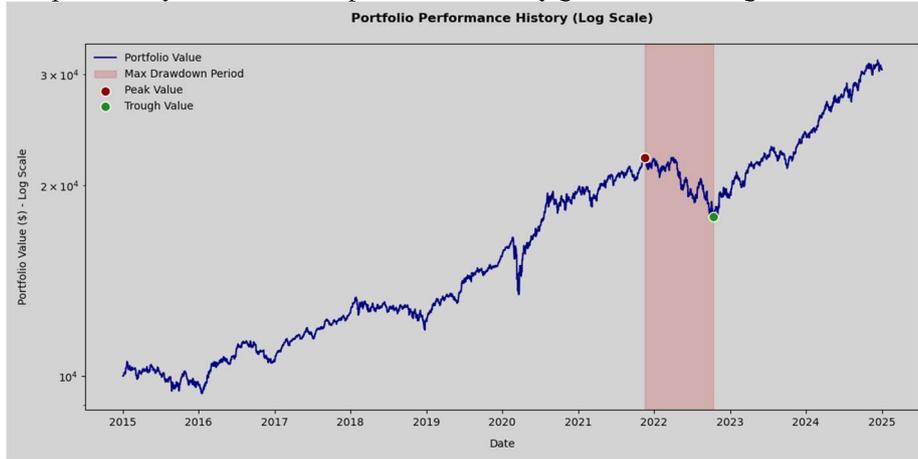

**Fig. 1.** Portfolio value in log scale over time with maximum drawdown period highlighted.

Consider a portfolio with US tickers weights: QQQ (20.1%), VOO (28.3%) and GLD (51.6%) in Fig. 1, the maximum drawdown is 19.43% from the first trading day of 2015 to the first trading day of 2025. By understanding and managing maximum drawdown, investors can better prepare for potential margin calls.

CAGR stands for Compound Annual Growth Rate. It is a metric used to measure the average annual growth rate of an investment, portfolio, or any other value that can fluctuate over time. Unlike simple average growth rates, CAGR considers the effect of compounding, providing a more accurate representation of growth over multiple periods. In the pursuit of optimizing a portfolio for maximum returns while minimizing the risk of a margin call, incorporating Compound Annual Growth Rate (CAGR) and Max Drawdown can provide a more comprehensive framework. With these in mind, MAR ratio is used.

The Managed Accounts Reports ratio (introduced by Leon Rose in 1978), is a performance metric that assesses a portfolio by balancing its returns against the associated risk. To compute the MAR ratio, divide the compound annual growth rate (CAGR) of a fund or strategy since its inception by its maximum drawdown (2).

$$\text{MAR Ratio} = \text{CAGR}/\text{Max Drawdown} \qquad (2)$$

the objective is to Maximize the MAR ratio of the portfolio optimization period, incorporating the MAR ratio into the portfolio optimization framework helps investors achieve a better balance between returns and risk, ultimately reducing the chances of a margin call and enhancing overall portfolio performance.

# 7      Example

## 7.1      Methodology

The U.S. equity market is one of the largest and most dynamic in the world, comprising thousands of publicly traded companies. With approximately 4,000 tickers available for investment, the sheer volume of options can be overwhelming for investors seeking to optimize their portfolios. To navigate this complexity and enhance the efficiency of portfolio optimization, a strategic approach involves implementing a pre-screening process as a constraint to filter desired stocks. Analyzing each stock individually for portfolio inclusion is impractical and inefficient. Therefore, a systematic method to narrow down the universe of potential investments is essential.

First step is to filter stocks based on market capitalization to focus on companies of a certain size. For instance, selecting the top 100 companies by market capitalization ensures a focus on large, established firms. Including large-cap stocks in a portfolio can provide stability and resilience, as these companies are better equipped to weather market volatility. Investing in large-cap companies can be a strategic move for investors seeking long-term growth, given their proven track record and established market positions. In the same sense, companies in the search space should be more than 10 years old

Investors generally strive to maximize returns while reducing the likelihood of a margin call. Striking this balance is essential for maintaining sustainable portfolio performance. A practical approach to meeting this goal is to optimize the MAR ratio, as it offers a thorough assessment of risk-adjusted returns.

To perform portfolio optimization with objective of maximizing the MAR ratio, Particle Swarm Optimization (PSO) of metaheuristics is used. It is particularly effective for optimizing complex, non-linear problems. PSO is computationally efficient and can converge to optimal solutions quickly, making it suitable for large-scale portfolio optimization problems. It can handle noisy and dynamic optimization landscapes, making it suitable for real-world portfolio optimization problems. It also can be scaled to handle large portfolios with numerous assets and complex constraints.

The portfolio optimization framework is now established, allowing portfolio optimization to proceed as planned. The results can be evaluated by comparing them to the metrics of The SPDR S&P 500 ETF Trust (SPY), a widely recognized benchmark for assessing the performance of U.S. equity portfolios. Top prominent companies are tracked across diverse sectors in the U.S. market.

## 7.2      Results

Portfolio optimization is carried out and below are the results generated.
US equities with top 100 capitalization and at least 10 years of history are as follow:
AAPL, NVDA, MSFT, AMZN, GOOGL, GOOG, META, TSLA, AVGO, BRK-B, BRK-A, WMT, LLY, JPM, V, MA, ORCL, UNH, COST, XOM, NFLX, HD, PG, JNJ, BAC, ABBV, CRM, TMUS, KO, CVX, WFC, CSCO, IBM, PM, ABT, MS, GE, MCD, AXP, ISRG, MRK, GS, TMO, NOW, ADBE, BX, DIS, PEP, QCOM, T, AMD, VZ, CAT, TXN, BKNG, SPGI, INTU, RTX, C, AMGN, BSX, PGR, UNP, BLK, SCHW,

DHR, SYK, PFE, LOW, NEE, TJX, BA, AMAT, ANET, CMCSA, HON, PANW, FI, DE, GILD, SBUX, ADP, KKR, COP, VRTX, PLD, MMC, MU, BMY, NKE, MELI, ADI, LRCX, INTC, KLAC, LMT, UPS, IBKR, WELL, ICE

**Table 1.** Optimization Results vs Benchmark

|  | Optimized Portfolio | Benchmark |
| --- | --- | --- |
| Period | 01/01/2015 to 01/01/2025 | |
| Rebalancing frequency | Annually | |
| Weights | LLY 68.2%, NVDA 31.8% | SPY 100% |
| CAGR | 50.64% | 13.01% |
| Maximum Drawdown | 21.52% | 33.72% |
| Sharpe Ratio | 1.18 | 0.38 |
| Sortino Ratio | 2.59 | 0.68 |
| CAGR/MaxDD Ratio | 2.35 | 0.39 |

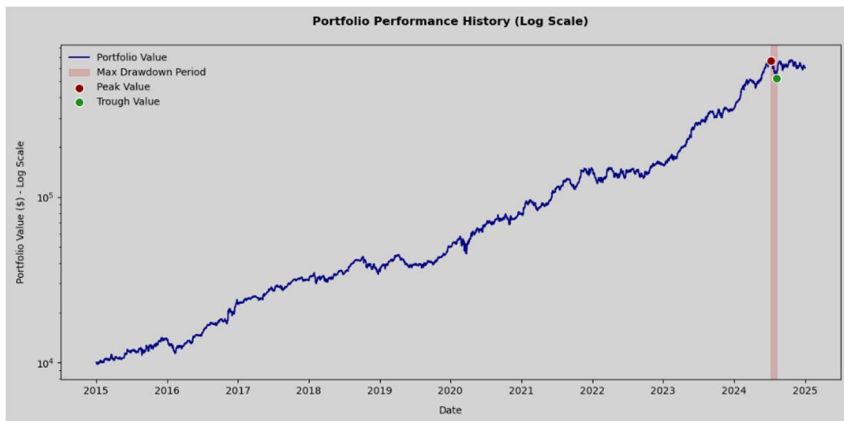

**Fig. 2.** Optimized Portfolio's value in log scale over time.

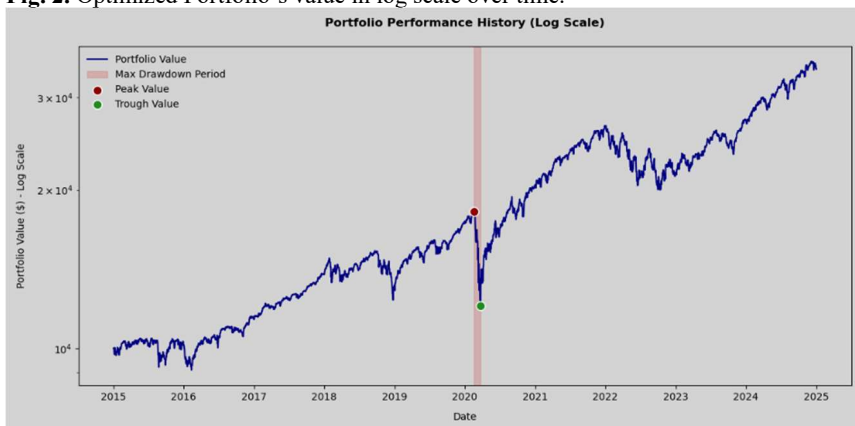

**Fig. 3.** SPY only portfolio's value in log scale over time.

The results of the optimization indicate that the max drawdown is effectively managed (2)(3), and the MAR ratio is favorable compared to the benchmark SPY (Table 1). The optimized portfolio exhibits a lower maximum drawdown compared to the benchmark SPY, indicating greater resilience during market downturns and periods of volatility. This results in a more stable investment experience. Additionally, the optimized portfolio achieves a higher MAR ratio than SPY, reflecting superior risk-adjusted returns. A portfolio with better MAR ratio generates more robust returns in comparison to down side variation, as quantified by its maximum drawdown. The controlled max drawdown indicates that the portfolio is more resilient to market fluctuations, reducing the likelihood of significant losses and margin calls.

## 8      Suggestions for future work

This paper presented a novel comprehensive framework for analyzing portfolio optimization using metaheuristics, specifically focusing on maximizing the MAR (Managed Accounts Reports) ratio and considering the max drawdown of the portfolio. Future work could explore the integration of dynamic margin requirements into the optimization framework. Margin requirements can fluctuate based on market conditions, volatility, and regulatory changes. Developing adaptive models that adjust portfolio allocations in response to changing margin requirements can enhance the framework's robustness. It is worth to investigate the impact of different margin call thresholds on portfolio performance and risk. By setting varying thresholds, researchers can identify optimal levels that balance the need for risk management with the potential for higher returns. The optimization framework could broaden to incorporate various performance metrics, like the Sortino ratio. This multi-objective strategy offers a more thorough assessment of portfolio performance and risk.

## 9      Conclusions

In conclusion, the optimized portfolio's performance, as evidenced by the controlled max drawdown and improved MAR ratio compared to the SPY, highlights the effectiveness of the portfolio optimization framework. This approach provides investors with a robust strategy for achieving high risk-adjusted returns while managing risk of margin trading efficiently. Further research may incorporate additional performance measurements which includes Sortino ratio or Sharpe ratio, to provide a deeper understanding of the portfolio's risk-adjusted returns.

**Disclosure of Interests.** The authors have no competing interests to declare that are relevant to the content of this article.